# Influence of KCl and HCl on a Laser Clad FeCrAl Alloy: In-Situ SEM and Controlled Environment High Temperature Corrosion

L. Reddy[1], M. Sattari[2], C. J. Davis[3], P. H. Shipway[1], M. Halvarsson[2], T. Hussain[1]

[1] Faculty of Engineering, University of Nottingham, Nottingham, NG7 2RD, UK

[2] Division of Materials Microstructure, Department of Physics, Chalmers university of Technology, Kemivägen 10, 41296, Gothenburg, Sweden

[3] Materials and Corrosion Team, Technology Centre, Uniper Technologies Limited, Ratcliffe on Soar, Nottingham, NG11 0EE, UK

# 1 Abstract

Increasing pressure on the power industry to reduce carbon emissions has led to increased research into the use of biomass feedstocks. This work investigates the effects of HCl and KCl, key species influencing biomass boiler corrosion, on a laser clad coating of the FeCrAl alloy Kanthal APMT. In-Situ SEM exposure of the coating at 450 °C for 1 h was performed to investigate the initial effects of KCl on the corrosion process. The same coatings were exposed to 250 h exposures in both an air environment and a HCl rich environment. The influence of KCl was investigated in both. Evidence of a slow growing aluminium oxide was observed. It was found that HCl allowed chlorine based corrosion to occur suggesting it can interact from the gas phase. It was also observed that the presence of both HCl and KCl reduced the mass gain compared to KCl in an air environment.

Keywords: In-Situ SEM, FeCrAl, APMT, High Temperature Corrosion, Chlorine

# 1. Introduction

As global legislation pressures countries to transition to more sustainable energy systems, there has been an increase in the uptake of biomass combustion as an alternative to conventional fossil fuels [1]. Whilst using biomass in a conventional fossil fuel boiler is in many ways similar to conventional fossil fuel combustion, it still comes with specific challenges. Biomass combustion can create a more aggressive combustion environment within a boiler, leading to increased material loss from boiler components and significant reductions in operational performance [2].

One of the techniques used to combat the aggressive high temperature corrosion environments within boilers is to use materials with higher corrosion resistance. Many of these materials are expensive, and as such it is not economically feasible to replace conventional materials in bulk. In these situations, a surface engineering solution or a corrosion resistant coating can be applied to the outside of a bulk material made from inexpensive, low performance alloy. One such technique for carrying out this coating process is laser cladding whereby the feedstock material is fed into a meltpool on the surface of the substrate generated by a laser. As the meltpool is moved across the surface a coating is built up.

Currently, materials used in heat exchanger components are mainly ferritic steels and the most commonly used coating materials for heat exchangers are corrosion resistant nickel based chromia forming alloys [3]. An alternative to these is high corrosion resistant iron-based alloys (FeCr, FeAl, FeCrAl, etc.) due to the lower cost and abundance of iron. The addition of small amounts of aluminium into the traditional FeCr alloys to form FeCrAl alloy has been shown to allow the formation of alumina at temperatures above 900 $^{\circ}$C [4]. Other work has shown that a thin alumina layer will form at temperatures of 600 $^{\circ}$C on a FeCrAl alloy, however this is most likely transient and so not stable and therefore non-protective [3]. The formation of stable alumina is beneficial as it forms dense, thermodynamically stable protective layer [5].

Salts play an important role in the corrosion performance of materials at high temperature. Corrosion performance can be influenced by these salts in both their solid and liquid phase. With typical meting temperatures in the rage of 500 $^{\circ}$C to 1000 $^{\circ}$C, these temperature ranges are the most studied. It is well established in the literature that chloride salts play a particularly important role in biomass induced fireside corrosion due to their high levels in many biomass feedstocks [6, 7]. In contrast to coal combustion, where typical deposits contain zinc and lead chlorides, most biomass combustion deposits are dominated by potassium chloride [8]. Hydrogen chloride is the most common chlorine containing species in biomass flue gas [8]. It has been shown that combustion of feedstocks can produce HCl [9]; however, this concentration is proposed to be increased by additional production as a by-product in mechanisms such as the formation of alkali chromates in the presence of alkali salts and water vapour [10]. The protective oxides which form on the surface of metals are typically dense, and prevent further diffusion of oxygen to the scale metal interface. This pacifies the surface and prevents further oxidation. HCl is able to diffuse through this dense protective oxide towards the metal scale interface. Once at this interface, where the partial pressure of oxygen is low, the hydrogen and chloride dissociate as metal chlorides become more energetically favourable. Once formed, the metal chlorides can diffuse back through the protective scale towards the scale gas interface. Here in the higher partial pressures of oxygen, the metal chlorides are oxidised, releasing the chlorine to diffuse back through the protective scale in the same cycle as the hydrogen chloride.

FeCrAl alloys offer an alternative to conventionally used low alloy steels. The addition of aluminium allows for the formation of alumina, as well as chromia scales. These scales are dense and thermodynamically stable. The current research into the effects of KCl and HCl on FeCrAl alloys is focused on two main areas: short scale, and long scale exposures. In short scale tests, it has been shown that pre-formed alumina scales, formed at 900 $^{o}$C for 1 h, provide protection against chlorine attack. Flaws in this protective alumina layer provide pathways for chlorine ingress and the eventual failure of the coating [10]. Without pre-oxidation, at 600 $^{o}$C, only an aluminium enrichment was detected at the scale coating interface, and potassium chloride was able to react with the FeCrAl alloy as well as $O_2$ and $H_2O$ present in the corrosion environment to form potassium chromate [3, 11]. The other area of research is at much longer time scales. Here the effects of chlorine are looked at in more complex environments and for much longer durations. Whilst playing an important protective role at short time scales, the concentration of chromium does not dictate the corrosion performance. However in longer scale tests, it is the concentration of chromium that will have the largest effect on corrosion performance, with aluminium concentration playing a less significant role [12]. At higher temperatures of 800 $^{o}$C in these more complex chlorine riche environments the level of aluminium became important as alumina began to form providing protection and reducing metal loss [13].

For much of the previous work discussed above, In-Situ Environmental Scanning Electron Microscopy (In-Situ ESEM) was used. In-Situ ESEM offers opportunities to investigate the progression of corrosion as it occurs. Environmental SEM chambers allow carful control of the gaseous environment surrounding a sample. This can be paired with real time imaging of the sample. The technique becomes more powerful when a heated stage is added into the chamber, allowing for high temperature corrosion tests to be imaged in real time. When looking at iron at 500 $^{o}$C and 700 $^{o}$C, in-situ measurements allowed for the structure of the initial iron oxides that grow on the surface to be investigated, providing insight into the earliest mechanisms to act [14]. The initial heating conditions of a sample can play an important role in its oxidation mechanisms. The use of in-site ESEM means that these initial heating stages can be observed [15]. It was found that KCl crystals melted at temperatures below 400 $^{o}$C, forming oxide shells in their presence. It was also found that a liquid film of these molten salts allowed the rapid transport of chlorine across the surface.

The existing literature investigates the performance of FeCrAl alloys at temperatures of 600 $^{o}$C or higher, and often in a pre-oxidised state. This work investigates a laser clad FeCrAl alloy on stainless steel at 450 $^{o}$C, a temperature matching the current operational temperatures of biomass fired boilers, and below much of the range currently explored in the literature. Despite this being a relatively small temperature change, it moves the temperature well away from the melting point of KCl, ensuring that interactions in the solid phase are investigated. Pre-oxidising components adds to their manufacturing coast and is unfeasible when retrofitting components. The laser cladding of corrosion resistant coatings is a cost effective alternative to bulk component manufacture. There is little work carried out in the literature looking at laser clad coatings, whose dilution and microstructure may influence the corrosion performance of the coating material. In this work, the effects of Potassium Chloride crystals and Hydrogen Chloride gas on the formation of alumina and other corrosion products on a laser clad FeCrAl alloy is investigated.

# 2. Experimental

## 2.1. Materials

Commercially available feedstock wire Kanthal APMT™ (Sandvick Heating Technology AB, Sweden) with composition (C < 0.08, Al = 5.0, Si < 0.7, Cr = 21.0, Mn < 0.4, Fe = 69.8 and Mo = 3.0 all in wt.%) was laser clad onto a stainless steel 304 substrate (C < 0.08, Cr < 20, Mn < 2, Ni < 12, N < 0.1, P < 0.045, Si < 0.75, S < 0.03 and Fe Bal. all in wt.%). A front fed wire feed setup was used to feed the 1.2 mm wire feedstock into the laser spot. A 2 KW ytterbium-doped fibre laser (IPG Photonics, Germany) with a 600 µm fibre was used to deposit the Kanthal APMT wire feedstock onto a 200 mm x 100 mm x 6 mm 304 stainless steel plate. A spot size of 4 mm was produced with a 20 mm defocus. A power of 1800 W was used with a bed traverse speed of 3.3 mm s$^{-1}$ and a wire feed rate of 6.6 mm s$^{-1}$. Each successive pass of the laser deposited at a step distance of 2.618 mm producing a 40% overlap coating. These parameters were selected from previous work on optimising coatings for corrosion performance [16]. This section of coated substrate was cut into 10 mm x 10 mm sections using a silicon carbide precision cut off disc. The top surface of each of these sections was ground sequentially with silicon carbide grinding papers and finally to a diamond finish to 1 µm.

X-Ray Diffraction (XRD) was carried out on the top surface of the sample to determine the phases present within it. The XRD was carried out on a D500 Diffractometer (Siemens AG, Germany) with a diffracted-beam monochromator and scintillation counter detector. The instrument was operated at 40 kV and 20 mA to generate Cu Kα radiation at a wavelength of 0.1540 nm. The XRD scans were performed in the range 30 ° ≤ 2θ ≤ 90 ° with a step size of 0.08 ° and a dwell time of 8 s. This was performed both pre and post exposure. Micrographs of the top surface of the coating were taken on an S-3400N Scanning Electron Microscope (SEM) (Hitachi High Technologies, IL) in Hi-Vac mode with 20 kV beam voltage, Working distance of 10 mm and spot size 4. Energy-Dispersive X-ray (EDX) spectroscopy was performed on the segregated areas to determine their composition. An 80 mm$^2$ spectrometer (Oxford Instruments, UK) was used to take these measurements and the analysis software INCA (Oxford Instruments, UK) to draw elemental information. This was again performed pre and post exposure.

## 2.2. Methodology

The samples were deposited with a saturated solution of KCl in water and industrial methylated spirit pipetted onto the surface. In-situ high temperature oxidation tests were carried out in an FEI Quanta 200 Field Emission Gun Environmental Scanning Electron Microscope (FEG ESEM) (Thermo Fisher Scientific, MA). The setup of the SEM is the same as that found in previous work [15]. The prepared samples were exposed in laboratory air at 1 Torr for 1 h. The samples were heated at 450 °C for the duration of the exposure using a heated stage attached to a temperature control unit. The individual KCl crystals had micrographs taken at regular intervals throughout the exposure is secondary electron mode.

Two test environments were used to carry out the high temperature controlled environment corrosion tests. The first environment was a circulating air box furnace, the second was a synthetic biomass flue gas containing 500 ppm HCl, 5 % $O_2$ and $N_2$ balance, performed in a horizontal tube furnace with an internal diameter of 70 mm. The stainless steel reactor used in the furnace was lined with high purity alumina. The samples prepared in 2.1 were placed in individual alumina crucibles in each environment. Exposures were conducted in both environments for 250 h at 450 °C, as well as for 1 h in air at 450 °C to match the in-situ ESEM exposure of the sample. Before exposure, the samples were degreased and cleaned using ultrasonic agitation in industrial methylated spirit. The samples were split evenly into two groups. One group had a saturated solution of KCl in water and industrial methylated spirit pipetted onto the surface, which was left to evaporate, leaving individual KCl crystals. The second group were left polished with no deposit. Samples from each of these groups were exposed in both test environments.

After surface characterisation, cross-sections were made of the samples. These cross-sections were cut, ground and polished using the same process as for the initial surface preparation, however non-aqueous lubricants were used to ensure chlorides were not removed. These sections were cold mounted to ensure the protection of any deposit formed and were again analysed with SEM and EDX using the same instrumentation to form a complete analysis of the samples.

# 3. Results

## 3.1. In-situ SEM exposure of APMT at 450 $^o$C

### 3.1.1. Without KCl deposit

Figure 1 (A) shows the exposed top surface of the APMT coating deposited with KCl. Figure 1 (B) shows a TEM section through a KCl crystal at a point where the obvious corrosion deposits resulting from the KCl crystal are not present. Two layers of protective platinum can be seen on top of the oxide. As oxygen has been excluded from the EDX measurements, all percentages represent cation percentage. At regions where the influence of the KCl crystal is not immediately obvious, a dual layer oxide can be seen to have formed on the surface. Moving from point A inwards, the chromium increases, from a minimum value at A of 9.9 % to a maximum of 43.1 %, well above the bulk composition. This is matched by a decrease in iron concentration over the same distance from its maximum at A of 82.6 % to a minimum of 43.9 %. This region represents an iron oxide with a thickness of approximately 10 nm, giving a total oxide thickness of 16 nm. In Figure 2, A high concentration of aluminium can be seen at the oxide coating interface, surgesting the presence of aluminium oxide. Beyond this region, across the boundary at B, there is a very rapid chromium depletion over a distance of approximately 6 nm. Here chromium concentration reaches 13.8 %. This is again matched by a change in iron as its concentration raises back to 74.5 % over the same region. As oxygen has been excluded from the EDX measurements, this transition region most probably represents the iron-chromium spinel. Moving inwards from point B we see a small region in Figure 2 below the boundary at B, approximately 15 nm thick where chromium returns to its nominal bulk value of between 21.6 ± 0.1 %. Iron also returns to its bulk value over this region of 67.7 ± 0.1 %.

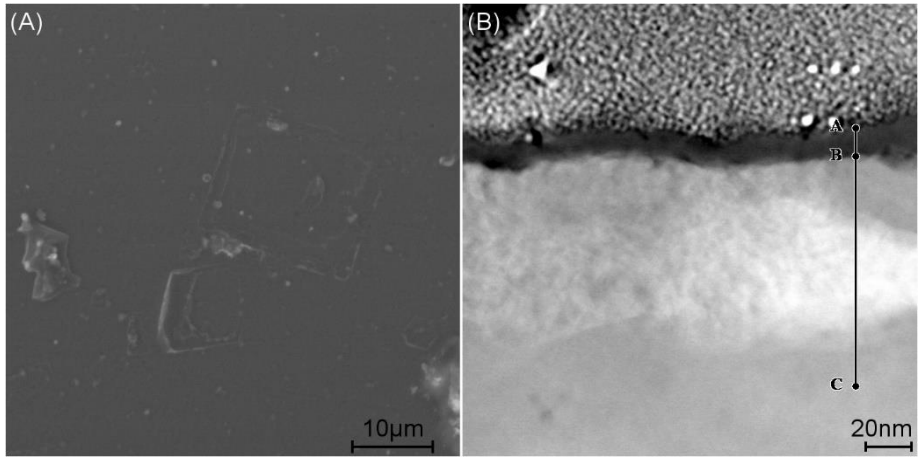

*Figure 1: (A) SE micrograph of top surface and (B) STEM-HAADF image of a FIB lift out cross section of the surface of the APMT coating exposed in laboratory air at 1 Torr at 450 °C. The section is taken through the site of a crystal shown in (A).*

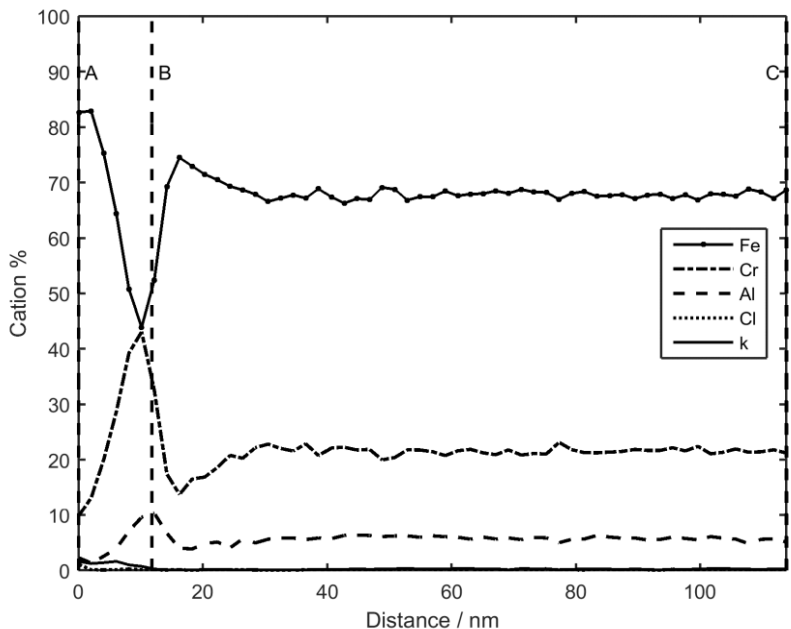

*Figure 2: Elemental distribution along the line shown in Figure 1. There is an outer iron oxide which decreases as an iron-chromium oxide grows between points A and B. An aluminium rich region can also be seen at the interface denoted by the line B.*

### 3.1.2. With KCl deposit

Figure 3 shows the evolution of a potassium chloride crystal during heating at 450 °C over the course of 1 h in the in-situ SEM setup. It can be seen that in the case of this crystal, the KCl is evaporated during the heating, beginning at around 30 mins. Regions of corrosion product can be seen beginning to form around the edge of the crystal as the KCl is consumed, and become especially evident after 40 min when the remaining KCl crystal has completely evaporated. Cross-section of this evaporated crystal can be seen in Figure 4. A thin layer of oxide can be seen above the surface of the coating between points A and B. This is shown to be more separated from the underlying coating than that of the sample taken far from the evaporated crystal shown in Figure 1. From Figure 5, it can be seen that a double layered oxide has again formed, similar to that seen in the sample where the influence of the KCl is not obvious. In the outermost region, at point A, an iron oxide forms with the chromium concentration dropping to 8.3 % and the iron content increasing to 83.6 %. Moving inwards, towards the coating oxide interface, most likely an iron-chromium spinel forms approximately 6 nm thick. This contains 46.6 % chromium and 30.3 % iron.

In the bulk material there is a depletion of both iron and chromium close to the coating oxide interface at B. In the approximately 4 nm before the boundary, iron is depleted from the bulk level to 61.0 %. The chromium depletes to a lesser extent but over a greater distance, dropping to 14.9 % over a distance of approximately 22 nm. The region between points B and C again represent the bulk material and have a composition of 68.4 ± 0.2 % iron and 21.6 ± 0.2 % chromium. Figure 5 also shows an aluminium enrichment at the coating oxide interface where the concentration is 10.0 % compared to its measured bulk value between 5.0 % and 6.1 %. Finally, the presence of the KCl crystal can be evidenced, in the peaking of chlorine and potassium in the iron-chromium spinel layer with concentrations of 10.0 % and 3.3 % respectively.

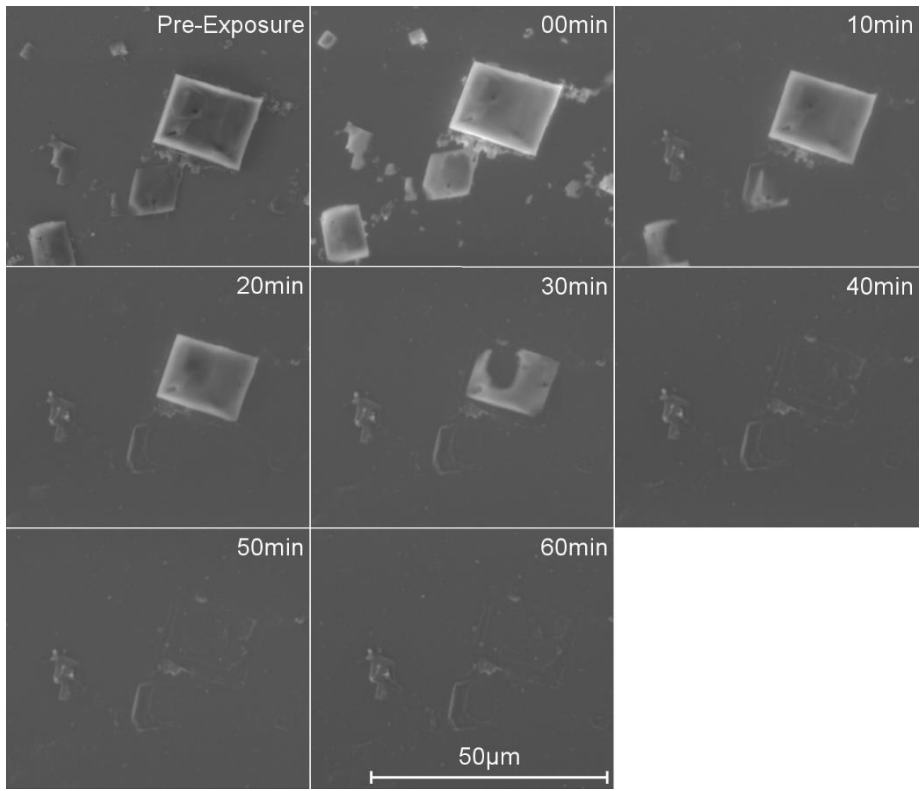

*Figure 3: Time evolution of a KCl crystal on the surface of APMT coating exposed in-situ in laboratory air at 1 Torr at 450 °C over a 1h timescale. The KCl crystal is consumed, however a thin deposit can be seen around its perimeter. All micrographs have the same scale.*

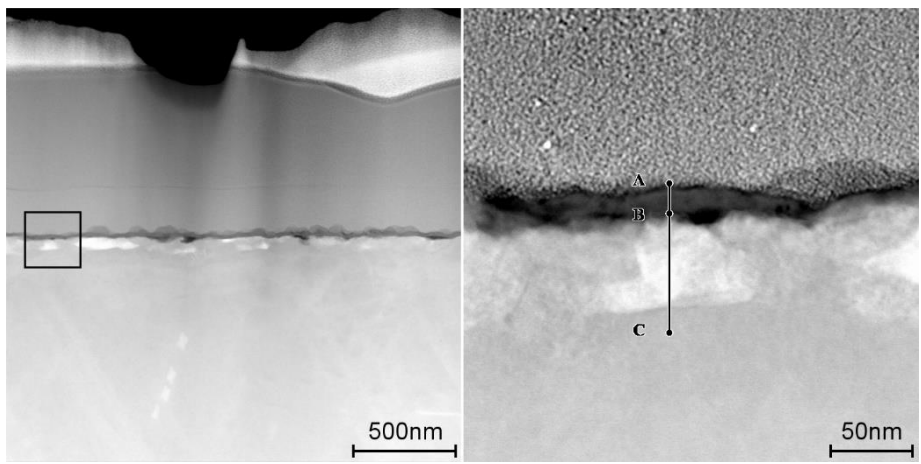

*Figure 4: STEM-HAADF images of a section of APMT coating exposed in laboratory air at 1 Torr at 450 °C. The section is taken close to the edge of a KCl crystal. In the left hand micrograph, platinum can be seen in the bright contrast and dark contrast region above the thin dark band of oxide. The micrograph on the right is a high magnification of the area marked by the box.*

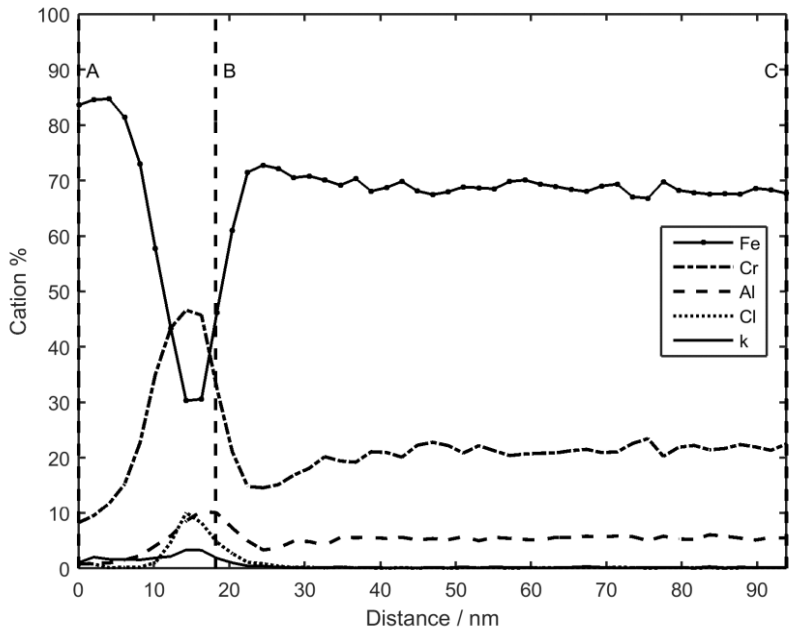

*Figure 5: Elemental distribution along the line shown in Figure 4. A double layered oxide is seen in the dark region with an outer iron rich oxide, and an inner mixed chromium iron oxide. A rise in chlorine can be seen matching with the increase in chromium in the same region.*

### 3.2. XRD phases

Figure 6 shows XRD scans of the coatings exposed in air and HCl for varying times without the presence of KCl. It can be seen in the spectra for the as-clad APMT where the peaks for the substrate lie. Whilst this phase could not be indexed accurately, it closely matches a number of iron-chromium and iron-aluminium mixed phases (PDF 03-065-4664, PDF 00-045-1203 respectively) with peak positions at 44.4 °, 64.6 ° and 81.7 ° and associated d-spacing of 2.04 Å, 1.44 Å and 1.18 Å respectively. As such it is suggested that this is an iron solid solution containing chromium and aluminium, as would be expected from APMT. After 1 h there were no oxide peaks found within the detection limit of the XRD. After 250 h in both air and HCl, peaks can be identified matching the crystal structure of haematite and chromia. It can also be seen that the presence of HCl increases the intensity of these peaks. Aside from these peaks there are no other phases visible from the XRD scans and as such, any other phases present under these conditions are expected to be very thin.

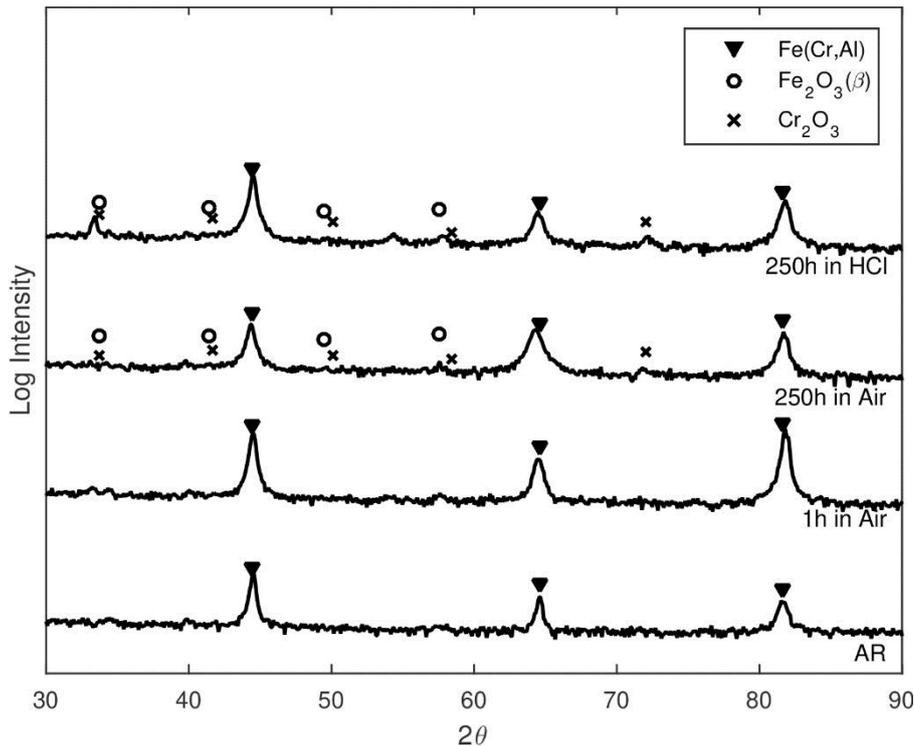

*Figure 6: XRD of APMT coating, as-coated and after 1 h and 250 h exposure at 450 °C in air and HCl. There are no obvious oxide phases present when compared with a pre-exposure sample after 1 h; however, very low intensity peaks occurring most notably at 58 ° and 72 ° appeared after 250 h, corresponding the presence of Fe$_2$O$_3$ and Cr$_2$O$_3$ which may be see in 3.1.1.*

Figure 7 shows XRD scans of the coating exposed in air and HCl for varying times with KCl. As observed in the samples exposed without the presence of KCl (as was observed in Figure 6), the same iron solid solution phase can be seen in all four scans. After 1 h in air in the presence of KCl, the β-hematite that was observed without KCl can be seen. This is accompanied by α-hematite which can be seen at all times and atmospheres when KCl is present. The other major phase noticed is potassium chloride. This can be seen in all of the samples with potassium chloride deposits and is never fully consumed. After 250 h in the presence of HCl the intensity of KCl in comparison to the other samples is lower, suggesting more of the KCl has been consumed. In the presence of air, the coating peaks cannot be observed. This could be due to a thick layer of KCl covering the surface and obscuring the substrate. There is an unindexed peak centred at a 2θ of 58 °. This peak appears in all samples in the presence of KCl, however a match could not be found.

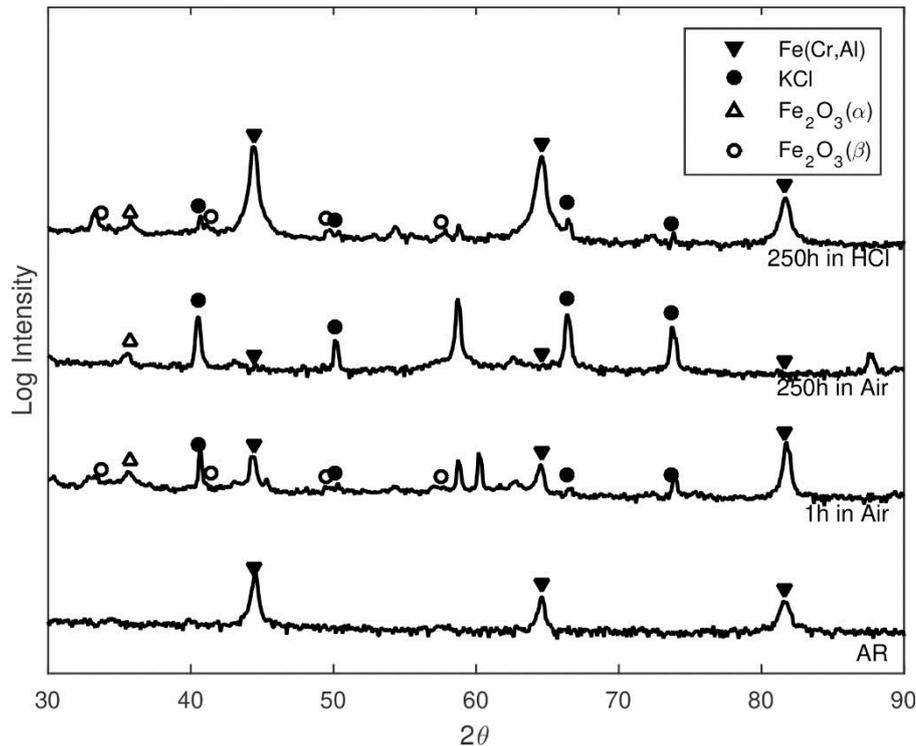

*Figure 7: XRD of APMT coating, as coated and after 1 h and 250 h exposure at 450 °C in air and HCl, with a KCl deposit. The APMT peaks are visible at all times except 250h, where the APMT is not visible.*

### 3.3. Long term exposure of APMT at 450 °C in air

#### 3.3.1. Without KCl deposit

After completing exposures in air for 1 h, long term exposures were carried out for 250 h. The top surface of the sample exposed without a KCl deposit can be seen in Figure 8. Figure 8 (A) shows a low magnification image of the top surface. Much like the surface of the sample exposed for 1 h, there are very few features that can be seen on the surface of the sample. The features that can be seen are shown in higher magnification in Figure 8 (B). Table 1 provides information on the composition of these phases. Whilst point 1 is iron and chromium deficient compared to the substrate measured at point 2, it is enriched in manganese suggesting a manganese oxide on the surface. This was not picked up in the XRD shown in Figure 6 however these oxides are very small.

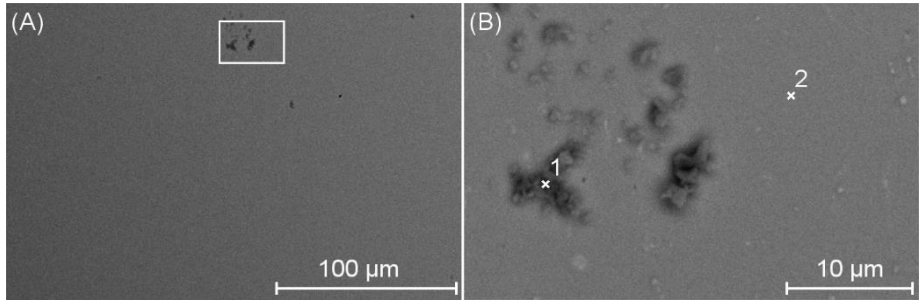

*Figure 8: BSE Micrograph of the oxides that have grown on the surface of APMT after exposure in air atmosphere for 250h at 450 ºC. In (A), isolated deposits of oxide can be seen on the surface. A high magnification of the area marked in (A) is shown in (B). All of the oxides in this field of view have a similar morphology.*

*Table 1: Composition of Figure 8*

|   | Fe | Cr | Al | Si | Ni | Mo | Mg |
|---|------|------|-----|-----|-----|-----|------|
| 2 | 57.3 | 18.1 | 4.2 | 3.4 | 2.9 | 2.9 | 11.3 |
| 4 | 69.0 | 21.0 | 3.2 | 0.4 | 3.3 | 3.2 |      |

### 3.3.2. With KCl deposit

When exposed for 250 h in the presence of KCl, the deposits which grow on the surface of the APMT, seen in Figure 9 are very different to those that are observed at shorter duration and without KCl. Figure 9 (A) shows a low magnification image of the surface. It can be seen that there is a continuous and varying covering over the entire surface of the coating. Higher magnification images show some of the complex phases present in higher magnification. Figure 9 (B) shows a thick, extruded tube like structure. These structures have a cracked layered surface and can be seen from Table 2 to be rich in chlorine and chromium. The point 1 is the underlying iron oxide layer, however it can be seen that this has been penetrated by chlorine in Table 2. The majority of the surface is covered with the oxide under point 2, another iron rich oxide again permeated with chlorine and potassium. Finally, point 3 shows the presence of potassium chromate.

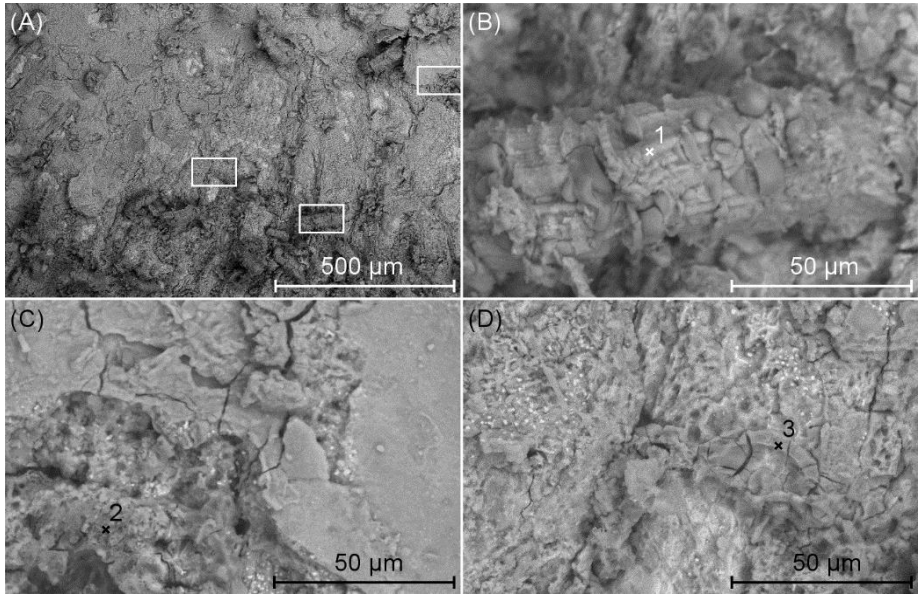

*Figure 9: BSE Micrograph of the oxides that have grown on the surface of APMT after exposure in air atmosphere for 250h at 450 °C in the presence of KCl. In (A), the continuous deposit across the surface can be seen. A high magnification of the areas marked in (A) are shown in (B), (C) and (D).*

*Table 2: Composition of Figure 9 showing EDX results from points 1 to 3 in wt.%*

|   | Fe | Cr | Al | Cl | Si | Ni | Mo | K |
|---|---|---|---|---|---|---|---|---|
| 1 | 14.3 | 33.7 | 12.0 | 25.5 |  | 1.9 | 3.7 | 9 |
| 2 | 65.9 | 11.0 | 2.1 | 6.5 | 0.7 | 4.6 | 3.3 | 5.8 |
| 3 | 23.0 | 32.9 | 5.1 | 11.8 | 0.8 | 2.8 | 7.5 | 16.2 |

### 3.4. Long term exposure of APMT at 450 °C with HCl

#### 3.4.1. Without KCl deposit

After investigating the behaviour of the coating for different exposure lengths in air, a synthetic flue gas rich in HCl was used for a 250 h test. This can be seen in Figure 10. At low magnification as seen in Figure 10 (A) a fairly regular deposit can be seen the entire surface, although it is not continuous. Figure 10 (B) shows a typical structure that makes up the majority of the deposits seen on the surface. This deposit is covered by point 1 in Table 3. It is very thin so shows mostly the underlying APMT coating however there is also chlorine present. Figure 10 (C) shows a plume like structure. These appear across the surface and look to originate from pores. These pores are not observed in the as-clad coating, nor after any of the other coated condition except when HCl is present. Figure 10 (D) shows another chlorine rich deposit on the surface, showing that the gaseous HCl can interact with the surface.

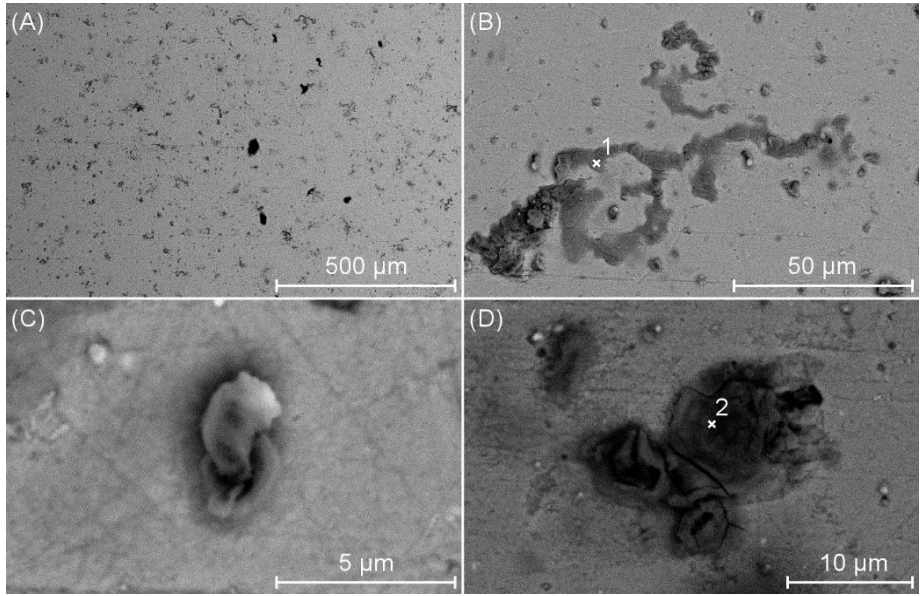

*Figure 10: BSE Micrograph of the oxides that have grown on the surface of APMT after exposure in HCl atmosphere for 250h at 450 °C. A thin, non-continuous oxide can be seen across the surface in (A). One region of oxide that have grown on the surface of APMT can be seen in (B). A plume like structure is shown in (C) which has grown out of the surface, seeming to originate from pores despite these not being visible on the unexposed surface. The other morphology of surface deposit not seen in (B) is shown in (D).*

*Table 3: Composition of Figure 10 EDX points 1 and 2 in wt.%.*

|   | Fe | Cr | Al | Cl | Si | Ni | Mo |
|---|---|---|---|---|---|---|---|
| 1 | 64.6 | 20.7 | 6.8 | 2.8 | 0.9 | 2.6 | 1.7 |
| 2 | 63.3 | 22.0 | 6.6 | 2.5 | 1.0 | 3.0 | 1.5 |

### 3.4.2. With KCl deposit

The final sample analysed was in a combination of both HCl gas and a KCl deposit. It can be seen in Figure 11 (A) that there is a complete covering of the sample with deposit. Figure 11 (B) shows the two components that make this up. There is a thin, continuous oxide that covers the entire surface. This is very thin and as such, is difficult to detect with EDX. The other feature on the surface is the chlorine rich tubes similar to those seen in Figure 9 (B). Finally, the plume like structures originating from pores that were seen in Figure 10 (C) are longer and better defined under these conditions.

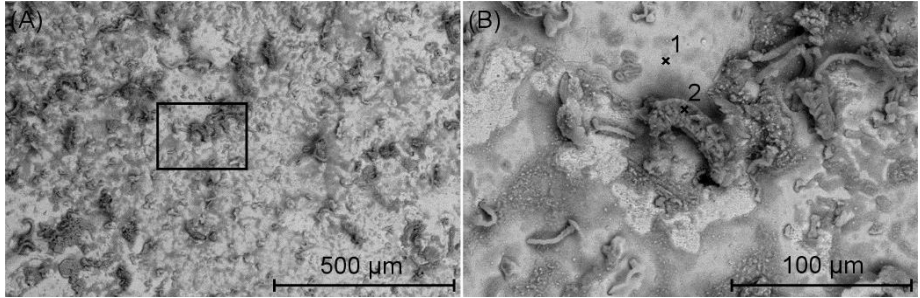

*Figure 11: BSE micrograph of the oxides that have grown on the surface of APMT after exposure in HCl atmosphere for 250h at 450 °C with a KCl deposit. A more comprehensive, continuous oxide than without the presence of KCl can be seen across the surface in (A). All of the phases present in the box in (A) can be seen in (B). They are all larger than those seen without the presence of KCl. In the presence of KCl the plumes that are shown in Figure 10 have grown larger and longer, however are still rooted in the pore like structures.*

*Table 4: Composition of Figure 11*

|   | Fe   | Cr   | Al  | Cl  | Si  | Ni  | Mo  | K   |
|---|------|------|-----|-----|-----|-----|-----|-----|
| 1 | 63.0 | 21.0 | 6.6 | 3.1 | 1.2 | 2.9 | 1.7 | 0.6 |
| 2 | 61.1 | 21.7 | 2.9 | 9.4 | 0.7 | 2.5 | 1.2 | 0.4 |

### 3.4.3. Mass change

Mass change was taken from the coated to the poste exposure samples. The greatest mass gain arises when KCl is present in both air and HCl environments. The addition of HCl seems to have little effect without the presence of KCl, however, when KCl is present, it is observed to have a protective effect. Whilst HCl and KCl individually exhibit corrosive effects on the coatings, combined, there effect is reduced in comparison to KCl alone.

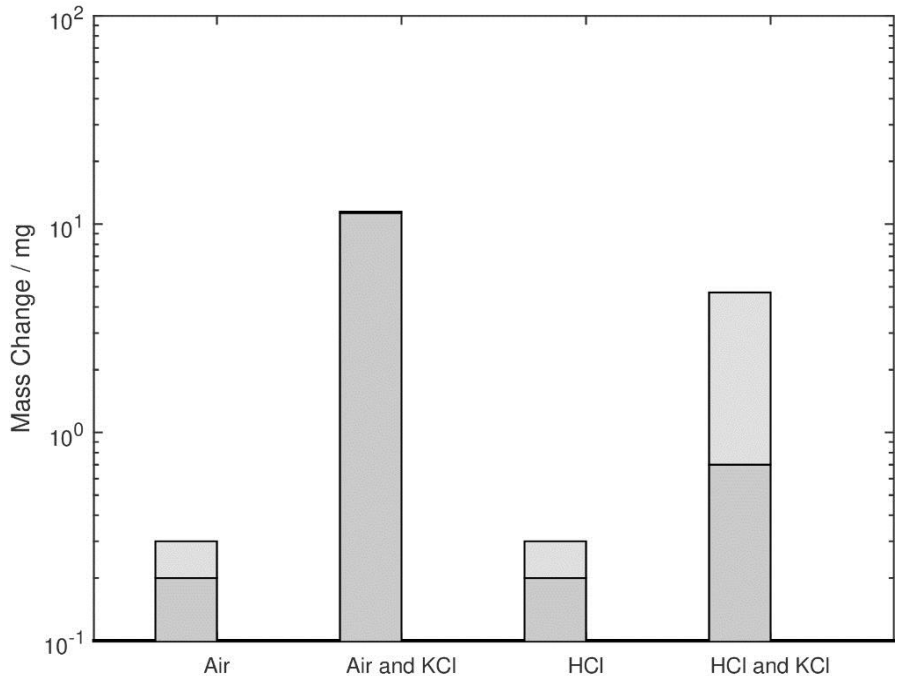

*Figure 12: Mass Change information for APMT coated samples in a range of environments and deposits. The light grey region represents the error in the measurement. In all environments, the presence of a KCl deposit increases mass gain. At 450 °C an HCl+KCl environment reduces the mass gain relative to a similar coating in air+KCl.*

# 4. Discussion

## 4.1. Oxidation in air and HCl

In the TEM cross-sections of the APMT clad exposed for 1 h in air, in a region where the influence of KCl is not obvious, there were three major phases identified at the coating-environment interface; an iron rich oxide, an iron-chromium mixed oxide and an aluminium enriched layer. In regions free from the presence of any KCl crystal on the top surface of the APMT clad, there was a chromium depletion zone observed within the coating inward of the innermost mixed iron-chromium oxide layer. This outward movement of chromium is what causes the observed relative decrease in iron. The formation of mixed iron and chromium oxides in stainless steel alloys is expected and well documented in the literature [17]. A relatively thin region of aluminium enrichment is observed suggesting that the formation of the iron and chromium oxides is rapid. The phase diagram of the iron-chromium-oxygen system shows that a continuous series of solid solutions of $Fe_2O_3$ and $Cr_2O_3$ form [17]. These can then form into spinels in solid solution with $Fe_3O_4$.

The final thing that can be noted from this in-situ exposure is the enrichment of aluminium at the initial coating-environment interface. The formation of aluminium oxide is expected to be very slow at temperatures of 450 °C. During early stages of oxidation, before the formation of thick corrosion products, alumina can form, however as fast growing iron and mixed oxide scales form, the oxygen partial pressure at the coating, scale substrate decreases, slowing the growth of alumina further, preventing a thick layer from froming. This structure of a mixed oxide with aluminium enrichment at the coating-oxide interface is observed elsewhere in the literature under similar conditions in the alloy Kanthal AF [11].

Long term exposures were carried out for 250 h in four different environments: air without KCl, air with KCl, HCl without KCl, and HCl with KCl. In the simplest environment, in air without KCl, the results observed were consistent with the short term exposures in that very little was observed on the surface within the limitations of SEM imaging. There is some evidence that manganese oxide formed on the surface. This is unexpected due to the minimal manganese in the coating (0.4 wt.%); however, this may be a region in which dilution of the substrate into the coating is high and as such the manganese concentration is higher in the coating than in the bulk.

Looking at the slightly more complex environment of HCl without KCl in long term exposures shows similar results after a 250 h exposure to that in air. The mass change of these two sets is very similar, initially suggesting that the presence of HCl does not affect corrosion; however, like in an air environment, the observable reaction products are very small, and as such the similarities in mass change could be due to measurements of such small amounts of reaction product; despite this chlorine was detected on the surface of the sample under SEM analysis. This could be particularly important when looking at the plume like structures that have grown from pores in the sample (Figure 10 (C)). These features could represent the oxidation of metal chlorides that have formed at the lower oxygen partial pressures present inside a pore [18]. These plumes were extremely small but could signify the initiation of the reaction products found in exposures with KCl discussed later.

## 4.2. KCl deposit induced corrosion

Looking at the same TEM cross sections (Figure 4) of the APMT after 1 h exposure in-situ, but in close vicinity to a KCl crystal, the same three major components are spotted as in the regions far from the KCl crystal: an iron rich oxide, an iron-chromium mixed oxide and an aluminium enriched layer. Additionally, a potassium chloride enrichment was observed at the original coating-environment interface. A chromium depletion zone was again observed within the coating after the innermost mixed oxide layer. When comparing the depth of this chromium depletion zone in the presence of KCl as opposed to areas far from the KCl crystals, it can be seen that both the depth of the depletion zone, and the overall thickness of the reaction products are increased in the presence of the KCl. This increased thickness is especially true of the iron oxide that has grown outermost. It is well established that the presence of chlorine can inhibit the growth of chromium oxides and as such could explain the increased thickness of the iron based outer oxide [3]. It can also be seen in Figure 4 (B) that there are voids that have formed along the oxide-coating interface. It is possible that these voids may have been formed by the removal of chromium in the form of chromium chlorides as has been seen in other work with iron-chromium-aluminium alloys [11]. This removal of chromium in the form of chlorides could also explain the increased depth of the chromium depletion zone despite the inhibited chromium oxide growth.

In the presence of KCl a much more extensive set of corrosion products are found on the APMT clad. These products are formed to a much greater extent than that which was observed without KCl. The formation of oxides exclusively at the sites of consumed KCl crystals is consistent with the results observed in the in-situ test, and supports the theory that KCl plays a role in increased mass gain in these samples. EDX of the regions around former KCl crystals show a decrease in chromium, when compared with other regions of the sample. This, paired with the expected increase in chlorine and potassium in these areas adds further credence to the theory that chlorine in selectively removing chromium in a volatile phase, and is consistent with the results obtained from in-situ measurements. The increased rate of reaction product formation in the furnace test compared to that performed in-situ is not unexpected and can be explained by the difference in oxygen partial pressure in each case, 4.46 kPa and 0.03 kPa respectively. The presence of potassium and chlorine is also expected from the results of shorter exposures; however, it shows that over longer time scales it has neither been completely consumed nor evaporated, and is instead still present on the surface of the sample. The cause of the formation of the tube like structures which are not observed at shorter exposures in-situ is unexplained.

With the presence of HCl instead of air, the most notable result is the reduced mass gain in the coating with KCl as can be seen in Figure 12. Whilst it can be seen that the mass gain is greater in both air and HCl when KCl is present, the mass gain in HCl with KCl is significantly lower than that observed in air with KCl. In an air environment, the only source of chlorine is from the dissociation of potassium and chlorine. The presence of HCl causes an increase in chlorine partial pressure which could slow the dissociation of potassium from chlorine and therefore cause the observed mass gain differences. It has been proposed that HCl can produce chlorine through the reaction shown in Equation 1 [19].

$$4HCl + O_2 = 2Cl_2 + 2H_2O \qquad\qquad\qquad\qquad\qquad\qquad\text{Equation 1}$$

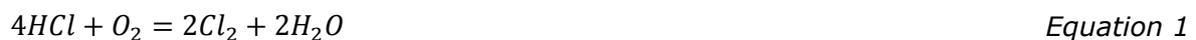

This is consistent with the chlorine detected in EDX on the sample exposed in HCl without a KCl deposit, suggesting that the HCl gas is providing some chlorine to the surface.

### 4.3. Generalised mechanism of corrosion of APMT

In the most simple environment of APMT clad in air, the formation of oxides can be explained by the following process [17]. Oxidation of aluminium to form $Al_2O_3$ happens very briefly until the oxygen partial pressure drops low enough that chromium and iron start to diffuse outwards. These form a $Cr_3O_4$ and $Fe_3O_4$ solid solution, which in time, form a $(Fe,Cr)_3O_4$ spinel in solid solution with the more abundant $Fe_3O_4$. This explains the multi layered oxides present and the formation of the spinel layer.

When HCl is added to the system, there is no reason to think that the above mentioned mechanism will not continue to take place. This is confirmed in the exposures seen where similar reaction products are formed. The formation of the plume like structures may be explained using the presence of HCl. It has been shown that Cl⁻, rather than HCl itself is the main aggressive species [20]. This can be formed through Equation 1. In low oxygen partial pressures, such as those found inside a pore, this newly formed chlorine ion can react with chromium or iron to form (Fe,Cr)Cl. These metal chlorides, as they diffuse towards the surface, are exposed to higher oxygen partial pressures where they undergo the following reaction according to Equation 2. The free chlorine ion is available to move back into the coating, allowing corrosion to continue as part of the chlorine cycle [21].

$$4MCl + 3O_2 = 4Cl^- + 2M_2O_3 \qquad \text{Equation 2}$$

When KCl is added to the system again, it gets more complex. Looking at KCl in an air environment, these mechanisms likely change again, which accounts for the vast differences in the corrosion products formed with and without KCl. As before, the initial stages of oxidation are the same, with the formation of $Cr_2O_3$. Once the $Cr_2O_3$ has formed, then it can react with the KCl and $O_2$ as through Equation 3. This liberates chlorine ion, which is then available to diffuse back through the deposit towards the metal coating where it can form metal chlorides, beginning the chlorine cycle discussed above and shown in Equation 2.

$$Cr_2O_3 + 4KCl + \frac{5}{2}O_2 = 2K_2CrO_4 + 4Cl^- \qquad \text{Equation 3}$$

In the most complex environment, a combination of all of these processes are in action. Initially, corrosion will follow the mechanism described above in Equation 3. The free chlorine ions originating from the KCl will play a role in the chlorine cycle, moving metal ions to the surface in the form of metal chlorides. Under normal conditions in air, Equation 2, these metal chlorides would be converted to metallic oxides at the surface, releasing chlorine to continue the cycle. This reaction will occur at a slower rate in the presence of HCl, due to the higher concentration of chlorine present at the surface already, formed from Equation 1. The net result of this is that potassium chromate is still formed; however, chromium chloride is not converted to chromium oxide, causing chromium chloride deposits to form, and reducing the formation of chromium oxide. This reduction in the formation chromium oxide accounts in part for the reduced mass gain observed when compared to APMT exposed in an air and KCl environment alone. The reduction in formation of chromium oxide through this mechanism is paired with the production of chromium chlorides. These chlorides are volatile, and can be removed at the gas oxide interface. This in turn contributes further to the reduced mas gain. It must be noted that this reduction in mass gain does not correlate to a reduction in corrosion rate, as chromium is lost though the volatile species produced.

This proposed mechanism is consistent with the results observed. In all cross-sections, an outer iron rich oxide is seen on top of an inner mixed spinel. The introduction of HCl shows the formation of a thicker oxide layer as well as evidence of potential potassium chromate forming from pores.

# 5. Conclusion

In this study, the corrosion mechanics of the iron-chromium-aluminium Kanthal APMT were investigated. The corrosion performance was tested in two environments. A short term corrosion test was carried out for 1 h at 450 $^{o}$C in an air environment. These tests were carried out in an in-situ ESEM. Longer term exposures were carried out in tube furnace for 250 h in air. These tests were also repeated in a HCl rich environment. These were also carried out at 450 $^{o}$C. All tests were carried out with and without the presence of KCl. The resulting corrosion products were analysed and the following conclusions made:

- Further evidence is found to support the role that chlorine plays in corrosion of APMT. Chlorine was found at the oxide coating interface after only 1 h when KCl was present.
- Aluminium is shown to migrate to the oxide coating interface suggesting that alumina may be slow growing in this region over longer time scales.
- Results obtained in-situ were able to be replicated in a furnace, observing similar levels of oxidation over the same time scales. This was replicated at longer time scales where oxide growth was greater but compositions were similar.
- In the presence of a hydrogen chloride environment the role of chlorine could still be seen suggesting that chlorine can interact with the surface from the gas phase as well as from the solid potassium chloride.

- The presence of hydrogen chloride provided protection to surfaces deposited with KCl. This is proposed to occur due to the breakdown of hydrogen chloride and oxygen into water and gaseous chlorine, slowing the dissociation of potassium and chlorine and hence limiting the rate of oxidation.

# 6. Acknowledgements

This work was supported by the Engineering and Physical Sciences Research Council [grant number EP/L016362/1]; in the form of an EngD studentship and industrial funding from Uniper Technologies Limited.

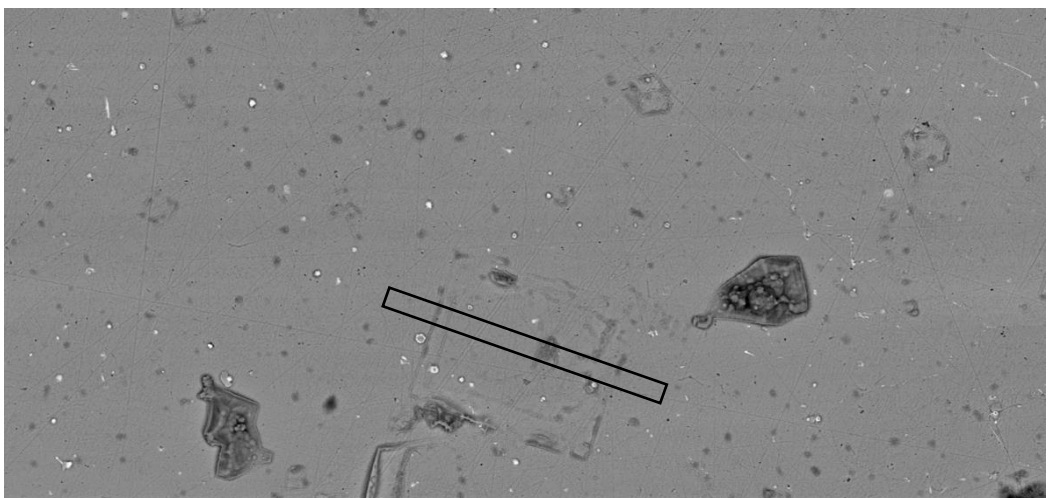

25